\documentclass[twocolumn,epjc3]{svjour3mod}          

\RequirePackage[T1]{fontenc}
\usepackage[normalem]{ulem}
\RequirePackage{graphicx}
\RequirePackage{mathptmx}      
\RequirePackage{flushend}
\RequirePackage[numbers,sort&compress]{natbib}
\RequirePackage[colorlinks,citecolor=blue,urlcolor=blue,linkcolor=blue]{hyperref}
\usepackage{latexsym,amsmath,amssymb,lmodern,float,url,bm,lineno}

\usepackage{flushend}

\usepackage{cellspace} 
\usepackage{ragged2e}
\apptocmd\thebibliography\justifying{}{}

\renewcommand{\Im}{\text{Im}\,}

\allowdisplaybreaks

\journalname{Eur. Phys. J. C}

\begin{document}

\title{Discovering the \texorpdfstring{\boldsymbol{$D_0^\ast(2100)$}}{D0*(2100)} in \texorpdfstring{\boldsymbol{$B$}}{B} semileptonic decays}

\author{M.-L. Du\thanksref{addr1}
        \and
        F.-K. Guo\thanksref{addr2,addr3,addr4}
        \and
        C. Hanhart\thanksref{addr5}
        \and
        F. Herren\thanksref{e1,addr6}
        \and
        B. Kubis\thanksref{addr7}
        \and
        \mbox{R. van Tonder}\thanksref{e2,addr8}
}

\thankstext{e1}{e-mail: florian.s.herren@gmail.com}
\thankstext{e2}{e-mail: raynette.vantonder@kit.edu}

\institute{School of Physics, University of Electronic Science and Technology of China, Chengdu 611731, China\label{addr1}
          \and
          Institute of Theoretical Physics, Chinese Academy of Sciences, Beijing 100190, China\label{addr2}
          \and
          School of Physical Sciences, University of Chinese Academy of Sciences, Beijing 100049, China\label{addr3}
          \and
          Southern Center for Nuclear-Science Theory (SCNT), Institute of Modern Physics, Chinese Academy of Sciences, \mbox{Huizhou 516000}, China\label{addr4}
          \and
          Institute for Advanced Simulation, Forschungszentrum J\"ulich, 52425 J\"ulich, Germany\label{addr5}
          \and
          Physics Institute, Universität Zürich, Winterthurerstrasse 190, 8057 Zürich, Switzerland\label{addr6}
          \and
          Helmholtz-Institut für Strahlen- und Kernphysik (Theorie) and Bethe Center for Theoretical Physics, Universität Bonn, \mbox{53115 Bonn}, Germany\label{addr7}
          \and
          Institut für Experimentelle Teilchenphysik, Karlsruhe Institute of Technology (KIT), 76131 Karlsruhe, Germany\label{addr8}
}


\maketitle
\begin{abstract}
The mass and width of the lightest scalar open-charm state listed in the Review of Particle Physics, the $D_0^\ast(2300)$, are in puzzling tension with predictions from unitarized chiral perturbation theory (UChPT) and lattice QCD, which favor a lighter state at around $2100$~MeV. However, to date, no direct experimental evidence for this lighter state exists.
In an effort to facilitate a direct observation, we introduce angular asymmetries of $B\rightarrow D \pi \ell \nu$ decays that allow for a direct extraction of the $D\pi$ S-wave phase shift and discuss a novel measurement strategy for the Belle II experiment. We conduct a sensitivity study, finding that the Belle II experiment can determine the pole location with sufficient precision to firmly establish the $D_0^\ast(2100)$ using the currently available data set.
We also investigate the possibility and necessary statistics of measuring the $D\pi$ isospin 1/2 scattering length with an accuracy sufficient to distinguish between the predictions from both UChPT and lattice QCD and the measurement by ALICE using femtoscopy.
\end{abstract}

\section{Introduction} 

Semileptonic decays of mesons are a very valuable source of information not only for tests of the Standard Model (SM), but also for hadron physics. For the former they provide stringent tests of lepton flavor universality involving semitauonic decays as well as theoretically clean avenues to perform precise measurements of Cabibbo--Kobayashi--Maskawa (CKM) matrix elements $|V_{ub}|$ and $|V_{cb}|$, allowing for sensitive tests of the SM by overconstraining the CKM unitarity triangle~\cite{Charles:2004jd,UTfit:2022hsi,HeavyFlavorAveragingGroupHFLAV:2024ctg}. For hadron physics, the arguably most valuable output is, thanks to Watson's final-state interaction theorem, phase shifts at low relative momenta also for particle pairs where direct scattering experiments are hindered by their short life times.
This allows for very stringent constraints on the properties of hadronic interactions and spectra. The prime example in this context is that a measurement of the low-energy pion--pion phase shifts from $K{\to}\pi\pi\ell\nu$~\cite{NA482:2010dug} not only provided the by far most precise experimental value for the isoscalar scalar scattering length, but also allowed for a significantly more precise determination of the pole of the $f_0(500)$~\cite{Caprini:2005zr, Garcia-Martin:2011nna, scalars:2024} 
that for a long time was subject of a
heated controversy in the community~\cite{Pelaez:2015qba}.

\begin{sloppypar}
Hadron spectroscopy in the charm sector has gained a significant boost at the beginning of this century when two states---$D_{s0}^*(2317)$~\cite{BaBar:2003oey} and $\chi_{c1}(3872)$, also known as $X(3872)$~\cite{Belle:2003nnu}---were found with properties severely at odds with the predictions of the simplest realization of the quark model that assigns mesons as quark--antiquark bound systems.
One way to pin down the structure of these potential exotics is to investigate their SU(3) flavor partners.
Here we focus on the family of singly-heavy scalar mesons, which not only hosts the $D_{s0}^*(2317)$ but also the $D_0^*(2300)$. In modern literature those states are typically viewed as quark--anti-quark bound states, dressed by two-hadron states~\cite{Yang:2021tvc,Ni:2023lvx,Zhang:2024usz},
 as compact tetraquarks with a 
diquark--anti-diquark substructure~\cite{Dmitrasinovic:2005gc,Maiani:2004vq,Maiani:2024quj},  or as hadronic molecules~\cite{Barnes:2003dj, Kolomeitsev:2003ac,Guo:2006fu,Guo:2017jvc}. 

Currently, the Review of Particle Physics lists the $D_0^*(2300)$ as the 
lightest open-charm state with scalar quantum numbers~\cite{ParticleDataGroup:2024cfk}, which appears to be rather heavy given that its strange partner state, the $D_{s0}^*(2317)$,
has basically the same mass. This 
mass proximity was interpreted as
an indication of a compact tetraquark structure for the two states~\cite{Dmitrasinovic:2005gc}.
In contrast, both UChPT and 
lattice gauge theory~\cite{Moir:2016srx,Gayer:2021xzv}, when extrapolated
to physical pion masses, predict
masses of approximately 2100~MeV, which are significantly lower~\cite{Kolomeitsev:2003ac,Guo:2006fu,Guo:2017jvc,Guo:2018gyd,Guo:2018kno,Guo:2018tjx,Lutz:2022enz}. 
The mass of 2300~MeV, extracted in experimental analyses using a single Breit--Wigner form, can be understood as arising from the interplay of two poles~\cite{Du:2017zvv}\footnote{In Ref.~\cite{Asokan:2022usm}, it is explained why the higher pole was overlooked in the pertinent lattice QCD studies.}
that appear inevitably from UChPT~\cite{Albaladejo:2016lbb}; for a more general discussion of 
two-pole structures, see Ref.~\cite{Meissner:2020khl}.
The heavier of the two poles was identified as a member of the flavor $[6]$ representation~\cite{Yeo:2024chk}, which is incompatible with an underlying $c\bar q$ core, but consistent with both a compact diquark--anti-diquark structure and a hadronic molecule. However, the former possibility was shown to be in conflict with the observed similarity of the scalar and axial-vector charmed meson spectra in lattice quantum chromo\-dynamics (QCD)~\cite{Guo:2025imr,Gregory:2025ium}.
\end{sloppypar}

Thus, lattice QCD and UChPT apparently agree 
that the lightest scalar charmed meson has a mass significantly lower than the nominal mass of 2300~MeV for the $D_0^*(2300)$, which is consistent with the molecular nature of all low-lying positive parity charmed mesons. 
However, a direct experimental support for this lower mass is still missing. While the $D\pi$ phase shifts extracted from the angular moments of $B^-\to D^+\pi^-\pi^-$ appear to be consistent with a pole at 2100~MeV~\cite{Du:2020pui}, no unambiguous conclusion can be drawn due to the complexity of the three-hadron final state, which leads to the dominance of a triangle diagram in the production process.

\begin{sloppypar}
What is clearly required is a direct experimental determination of both the low-energy $D\pi$ scattering parameters and 
the corresponding extraction of the pole location of the $D_0^*$. The isospin-1/2 $D\pi$ scattering length in leading-order
chiral perturbation theory is $\mu/(4\pi F_\pi^2) \approx 0.24$~fm~\cite{Guo:2009ct},\footnote{We use the particle physics sign convention, where a positive scattering length signifies an attractive interaction without a bound state.} with $\mu$ being the $D\pi$ reduced mass and $F_\pi$ the pion decay constant. This value
is enhanced to 0.4~fm by unitarization~\cite{Guo:2009ct,Liu:2012zya}, which simultaneously generates the
pole at 2100~MeV and is thus consistent with the phenomenology described above. The isospin-3/2 channel is found to have a scattering length of $-0.12$~fm, indicating weak repulsion, which remains essentially unchanged under unitarization.
In contrast, the ALICE Collaboration used the femtoscopy method
to determine the $D\pi$ scattering
lengths in the isospin-1/2 and 3/2 channels to be as small as $0.02(3)(1)$~fm and $0.01(2)(1)$~fm, respectively~\cite{ALICE:2024bhk}, values that are clearly at odds with those reported above. What is urgently needed is an independent, direct measurement of the isospin-1/2 $D\pi$
scattering length, not only to provide valuable information about the
low-lying $D$-meson spectrum, but also to 
better understand the accuracy of scattering length extraction from femtoscopy.\footnote{For a recent critical comment on the femtoscopy method, see Ref.~\cite{Epelbaum:2025aan}.}
In this work, we provide the foundations for such a measurement. In particular, we demonstrate to what extent existing and expected data will constrain our understanding of the low-energy regime of $D\pi$ scattering.
\end{sloppypar}

The remainder of this paper is structured as follows. In Sect.~\ref{sec:FFs}, we discuss the form factor decomposition for $B{\rightarrow}D\pi\ell\nu$ decays that is necessary for generating pseudo-data for this work and for future data analyses. In Sect.~\ref{sec:Obs}, we introduce observables directly related to the scattering phases. These observables allow us to extract both the pole location and the scattering length, for which we perform a sensitivity study in Sect.~\ref{sec:Study}. We conclude in Sect.~\ref{sec:Conc}.

\section{Description of \texorpdfstring{$B\rightarrow D\pi\ell\nu$}{B→Dπlν} decays}\label{sec:FFs}

The matrix elements entering semileptonic $B\rightarrow D\pi\ell\nu$ decays can be decomposed as~\cite{Gustafson:2023lrz,Herren:2025cwv}
\begin{align}
    &\left\langle D\pi |V^\mu|B\right\rangle = i\epsilon^\mu_{\nu\rho\sigma}p_{D\pi}^\rho p_B^\sigma \sum_{l > 0} L^{(l),\nu}g_l(q^2,M_{D\pi}^2)~,\nonumber\\
    &\left\langle D\pi |A^\mu|B\right\rangle =
    P^\mu f_+(q^2,M_{D\pi}^2) + \sum_{l > 0}\frac{1}{2}T_l^\mu f_l(q^2,M_{D\pi}^2)\nonumber\\
    &+P^\mu\frac{M_{D\pi}(M_B^2-M^2_{D\pi})}{\lambda_B}\sum_{l > 0} L^{(l),\nu}q_\nu\, \mathcal{F}_{1,l}(q^2,M_{D\pi}^2)\nonumber\\ &+ \dots ~,\label{eq:ffs1}
\end{align}
where we neglected two form factors only relevant to the case of massive leptons.
Here $p_\pi$, $p_D$, and $p_B$ are the pion, $D$-meson, and $B$-meson momenta, respectively. The momentum transfer to the leptonic system is given by $q^\mu = p_B^\mu-p_D^\mu-p_\pi^\mu$, $p_{D\pi} = p_D + p_\pi$, and $M_{D\pi}^2 = p_{D\pi}^2$.
The two vectors appearing in the decomposition are given by
\begin{align}
    T_l^\mu &= L^{(l),\mu}+\frac{4}{\lambda_B}\left[(p_B\cdot p_{D\pi})q^\mu - (p_{D\pi}\cdot q)p_B^\mu\right]L^{(l),\nu}q_\nu ~,\nonumber\\
    P^\mu &= (p_B + p_{D\pi})^\mu - \frac{M^2_B-M^2_{D\pi}}{q^2}q^\mu~.
\end{align}
The vector $L^{(l)}$ encodes the angular dependence of the final-state two-meson system and fulfills
\begin{align}
  L^{(l)}_\mu q^\mu  &= \frac{\sqrt{\lambda_B \lambda_{D\pi}}}{4M_B M^2_{D\pi}} P_l(\cos\theta) ~, &
    L_\mu^{(l)}p^\mu_{D\pi} &= 0~.
\end{align}
where the $P_l$ are the Legendre polynomials and $\theta$ the $D$-meson helicity angle.
The kinematic factors are given by
\begin{align}
    \lambda_B = \lambda(M_B^2,M_{D\pi}^2,q^2)~, \quad
    \lambda_{D\pi} = \lambda(M_{D\pi}^2,M_D^2,M_\pi^2)~,
\end{align}
with $\lambda(x,y,z)$ being the K\"all\'en function. Neglecting D waves and higher, four form factors enter: $f_+$ is the S-wave form factor, while $g_1$, $f_1$, and $\mathcal{F}_{1,1}$ are the three P-wave form factors.

In the following, we discuss the dominant $M_{D\pi}$-dependence of the form factors. Their $q^2$-dependence is treated through a $z$-expansion and unitarity bounds as in Ref.~\cite{Gustafson:2023lrz}.

\subsection{S wave}
\label{subsec:swave}
The goal of this work is to identify ideal observables
and to estimate the necessary luminosities to achieve sufficient extraction accuracy for the $D\pi$ isospin 1/2 scattering length and the $D_0^*$ pole to address the physics issues raised in the introduction. 
We thus need to generate pseudo-data based on a well motivated $D\pi$ amplitude. 

To describe the $D\pi$ S-wave lineshape in Ref.~\cite{Gustafson:2023lrz}, a coupled-channel approach based on the $D\pi/D\eta/D_s\bar{K}$ $T$-matrix from Ref.~\cite{Liu:2012zya} was employed. However, the uncertainty in the lineshape obtained was sizeable, as the relative normalization of the three channels was left floating in the fit to Belle data.
In the following, we improve upon this treatment by relating the three independent form factors.

At leading order (LO) of chiral perturbation theory, i.e., $\mathcal{O}(p)$ with $p$ denoting a general small momentum scale, the $D(p_1)\pi(p_2)\to D(p_3)\pi(p_4)$ scattering amplitude as given by the Weinberg--Tomozawa term is proportional to $(s-u)$, with $s=(p_1+ p_2)^2$ and $u=(p_1-p_4)^2$.
The leading contribution of its S-wave projection is given by $4 M_DE_\pi$,
where $E_{\pi} = (s + M_\pi^2 - M_D^2)/(2\sqrt{s})$ is the pion energy in the $D\pi$ center-of-mass (c.m.) frame.
Accordingly, at LO of the chiral expansion, the amplitude has an Adler zero located at $E_\pi = 0$, i.e., $s=M_D^2-M_\pi^2$. We have verified numerically that the next-to-leading order (NLO) correction to this approximation is less than 1 per mille, with the low-energy constants in the NLO chiral Lagrangian determined in Ref.~\cite{Liu:2012zya}. 
Such an Adler zero also appears in the production amplitude of the S-wave $D\pi$ system, and it is shown in Ref.~\cite{Du:2019oki} that neglecting the chiral behavior could lead to a sizeable overestimation of the $D_0^*$ resonance mass.
It is important to note that
the presence of the Adler zero, whose appearance is a necessary consequence of the chiral structure of QCD, 
also needs to be acknowledged when extracting physics 
parameters from the (pseudo)-data.

First, we observe that the three-channel generalization of the S-wave form factor $f_+$ obeys
\begin{align} 
\Im f_{+,i} = \sum_k T^\ast_{ik}\rho_k f_{+,k},
\end{align}
where the index $i=1,2,3$ represents the $D\pi$, $D\eta$,\footnote{Here, $\eta$ refers to the octet component of the physical eta meson. Singlet--octet mixing effects only enter at higher orders in the chiral expansion and, given the small influence of the $D\eta$ channel on the low-energy $D\pi $ system, will not affect the following discussion.} and $D_s \bar K$ channels, respectively, and
 $\rho_k(s) = {q_k}/(8\pi\sqrt{s})$, with
 $q_k$ for the $D_{(s)}\Phi$ ($\Phi=\pi,\eta,\bar K$) relative momentum, denotes the
 phase space of channel $k$. Without left-hand cuts, the solution reads~\cite{Au:1986vs}
\begin{align}\label{eq:sol:fi}
f_{+,i} = T_{ij}P_j(q^2)~.
\end{align}
For this study we drop a possible $s$ dependence of
the $P_j$, but allow for a $q^2$ dependence. Taking into account non-factorizable contributions is possible within the more involved parameterization developed in Ref.~\cite{Herren:2025cwv}, but would require double-differential experimental measurements in $q^2$ and $M_{D\pi}$. However, such contributions will cancel in the observables constructed in Sec.~\ref{sec:Obs}.

\begin{figure}[t]
\begin{center}
\includegraphics[width=0.4\textwidth]{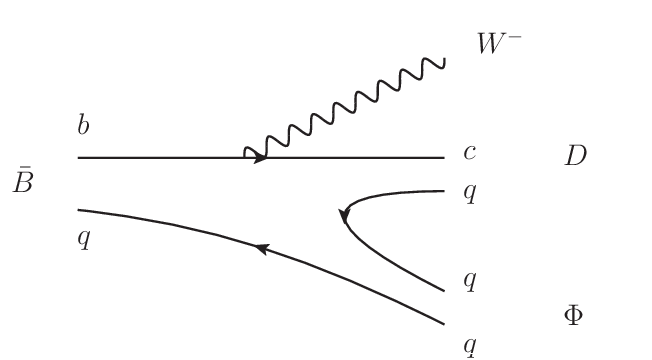}
\end{center}
\caption{Diagram for the semileptonic decay $B\to D\pi/D\eta/D_s\bar{K}$ at the quark level.}
\label{fig:feyndiag:quark} 
\end{figure}

The $P_j$'s can be related to each other when $SU(3)$ symmetry is implemented. The Feynman diagram for the semileptonic decay $B\to D\pi/D\eta/D_s\bar{K}$ at the quark level is shown in Fig.~\ref{fig:feyndiag:quark}.

In the low-energy region of $D\pi$, the effective Lagrangian can be expanded as 
\begin{align}
\mathcal{L}_\text{eff} = g_\text{eff} (\bar{l}\gamma^\mu(1-\gamma_5)\nu) B u_\mu D^\dagger + \dots ~.\label{eq:effLag}
\end{align}
Here $B=(B^-,\bar{B}^0,\bar{B}_s^0)$, $D=(D^0,D^+,D_s^+)$, and $u_\mu = i(u^\dag \partial_\mu u-u \partial_\mu u^\dag)$ with $ u =\exp(i\Phi/(\sqrt{2}F_0))$, $\Phi$ is the Goldstone boson matrix, and $F_0$ the pseudoscalar decay constant in the chiral limit.

At LO, a relation among the $P_j$'s can be read off from the Lagrangian~\eqref{eq:effLag}. 
For example, at tree level, the production amplitudes for $B\to D\Phi$ have the ratios
\begin{align}
A(B^-\to D^0\pi^0):A(B^-\to D^+\pi^-):A(B^-\to D^0\eta)\nonumber\\:A(B^-\to D^+_sK^-) = \frac{1}{\sqrt{2}} : 1: \frac{1}{\sqrt{6}}:1,
\end{align}
or
\begin{align}
A(B^- \to D\pi (I=1/2)):A(B^-\to D\eta  (I=1/2))\nonumber\\:A(B^-\to D_s\bar{K}  (I=1/2)) = \sqrt{\frac32}: \frac{1}{\sqrt{6}}:1.
\end{align}
With these, the
solution of Eq.~\eqref{eq:sol:fi} reads
\begin{align}
f_{D^+\pi^-}(q^2,s)&=P(q^2)\Bigg(\sqrt{\frac32}T^{I=1/2}_{D\pi\to D\pi}(s)\nonumber\\ &+ \frac{1}{\sqrt{6}} T^{I=1/2}_{D\eta\to D\pi}(s) + T^{I=\frac12}_{D_s\bar{K}\to D\pi}(s)\Bigg).
\end{align}
Consequently, only one unknown form factor remains. Since neither lattice calculations nor experimental data for this form factor exists, we treat its $q^2$-dependence following Ref.~\cite{Gustafson:2023lrz}, capturing the dominant features at high $q^2$ due to subthreshold axial-vector $B_c$-resonances, but leading to a sizeable uncertainty at low $q^2$.

\subsection{P wave}
\begin{sloppypar}
Due to the narrow width of the $D^\ast$ resonance, we approximate the P-wave form factors as in Ref.~\cite{Gustafson:2023lrz}:
\begin{align}
    \mathcal{F}_{1,1}(q^2,M_{D\pi}^2) &\approx \hat{\mathcal{F}}_{1,1}(q^2)h(M_{D\pi}^2)~, \nonumber\\ f_{1}(q^2,M_{D\pi}^2) &\approx \hat{f}_{1}(q^2)h(M_{D\pi}^2)~, \nonumber\\ g_{1}(q^2,M_{D\pi}^2) &\approx \hat{g}_{1}(q^2)h(M_{D\pi}^2)~.
\end{align}
Here $h$ encodes the $D^\ast$ line shape, which we parameterize by a relativistic Breit--Wigner form,
\begin{align}
    h(M_{D\pi}^2) = \frac{g_1\,F^{(1)}(M_{D\pi}^2,q_0)}{(M_{D\pi}^2-M_{R,1}^2) + i M_{R,1} \Gamma_R(M_{D\pi}^2)}~.
\end{align}
The line shape is parameterized through the $D^\ast$ mass $M_{R,1} = M_{D^\ast}$, the $D^\ast D\pi^\pm$ coupling constant $g_1 = 4.877~\text{GeV}$, determined from the $D^\ast$ width and branching ratios \cite{ParticleDataGroup:2024cfk}, and the P-wave Blatt--Weisskopf damping factor $F^{(1)}$.

In contrast to Ref.~\cite{Gustafson:2023lrz}, we follow Ref.~\cite{Hanhart:2023fud} to express the energy-dependent width by Chew--Mandelstam functions.
To this end, for decays into to unequal mass pseudoscalar mesons, we need to evaluate the dispersive integral
\begin{align}
    \Sigma_l(s + i\epsilon) = \frac{s - s_+}{\pi}\int_{s_+}^\infty\frac{\rho(s')n^2_l(s',q_0)}{(s'-s_+)(s'-s-i\epsilon)}\mathrm{d}s' ~,
\end{align}
where $s_\pm = (m_1 \pm m_2)^2$, $q_0 = 0.5~\text{GeV}$ is a fixed momentum scale, and
\begin{align}
        n_l(s, q_0) &= \left(\frac{q_k}{q_0}\right)^l F^{(l)}(q_k/q_0)~,\nonumber\\
        q_k(s) &= \frac{\sqrt{(s_+-s)(s_--s)}}{2\sqrt{s}}~.
\end{align}
While we do not need the $l=0$ case, it provides us with the only non-trivial integral we need to evaluate:
\begin{align}
        \Sigma_0(s + i\epsilon) &= \frac{s - s_+}{16\pi^2}\int_{s_+}^\infty\frac{\sqrt{(s_+-s')(s_--s')}}{s'(s'-s_+)(s'-s-i\epsilon)}\mathrm{d}s' \nonumber\\ &\equiv \frac{s - s_+}{16\pi^2}I(s)~.\label{eq::sig0}
\end{align}
The closed-form solution is given by
\begin{align}
    I(s) = \frac{2}{s(s_+ -s)}\Bigg[\sqrt{(s-s_+)(s_- -s)}\arccos{\sqrt{\frac{s-s_+}{s_- -s_+}}}\nonumber\\-i\sqrt{s_+ s_-}\left(1-\frac{s}{s_+}\right)\arccos{\sqrt{\frac{s_+}{s_+-s_-}}}\Bigg]
\end{align}
which reduces to the expression in Ref.~\cite{Hanhart:2023fud} in the equal-mass limit ($s_- = 0$). The calculation of the $l=1$ case can be significantly simplified by observing
\begin{align}
    \left(\frac{q_k}{q_0}\right)^2 F^{(1)}(q_k/q_0) = \frac{q_k^2}{q_0^2 + q_k^2} = \frac{(s-s_+)(s-s_-)}{(s-s_1)(s-s_2)}
\end{align}
with
\begin{align}
    s_{1/2} = \frac{1}{2}\left(s_+  + s_- - s_0 \pm \sqrt{(s_+  + s_- - s_0)^2 - 4 s_+ s_-} \right)
\end{align}
and $s_0 = 4 q_0^2$. A partial-fraction decomposition of the resulting integrand allows one to express $\Sigma_1$ 
as a sum of terms containing the $l = 0$ integral $I(s)$,
\begin{align}
    \Sigma_1(s) = \frac{s - s_+}{16\pi^2}&\Bigg[I(s)n_1^2(s,q_0) \nonumber\\ & + I(s_1)\frac{(s_1-s_+)(s_1-s_-)}{(s_1-s)(s_1-s_2)}\nonumber\\ &+ I(s_2)\frac{(s_2-s_+)(s_2-s_-)}{(s_2-s)(s_2-s_1)}\Bigg]~.
\end{align}
Since we do not include any other resonances or background terms in the P-wave, we can simply express the energy-dependent width together with the respective dispersive corrections as
\begin{align}
    i\Gamma(s) &= i\Gamma_{D^\ast\rightarrow D\gamma} + \frac{1}{M_{D^\ast}}\Bigg[\frac{g_1^2}{2}\Sigma^{D^0\pi^0}_1(s)\nonumber\\ &+ g_1^2\Big(\Sigma^{D^+\pi^-}_1(s)-\Sigma^{D^+\pi^-}_1(M^2_{D^\ast})\Big)\Bigg]~.
\end{align}
\end{sloppypar}

Following Ref.~\cite{Gustafson:2023lrz}, we fix the $q^2$-dependence and normalization of the form factors by integrating over $M_{D\pi}^2$ and equating to the lattice QCD calculation of the form factors by the Fermilab/MILC Collaboration~\cite{FermilabLattice:2021cdg}:
\begin{align}
    &\int\mathrm{d}M_{D\pi}^2 \frac{\mathrm{d}^5 \Gamma}{\mathrm{d}M_{D\pi}^2\mathrm{d}q^2\mathrm{d}\cos\theta\mathrm{d}\cos\theta_l\mathrm{d}\chi}\nonumber\\ &= \mathrm{Br}(D^\ast \rightarrow D \pi)\frac{\mathrm{d}^4 \Gamma}{\mathrm{d}q^2\mathrm{d}\cos\theta\mathrm{d}\cos\theta_l\mathrm{d}\chi}\Bigg|_\text{FNAL/MILC}~,
\end{align}
where $\theta_l$ and $\chi$ are the helicity angle of the charged lepton and the angle between the leptonic and hadronic decay planes, respectively.

\section{Directly measuring phases in semileptonic decays}\label{sec:Obs} 
\begin{sloppypar}
The rich angular structure of semileptonic four-body decays allows for the construction of five different observables sensitive to the relative phase between S and P waves. One is given by the forward--backward asymmetry in the helicity angle of the $D$-meson $\theta$, while the other four arise from contributions to the differential decay rate that vanish upon integration over the helicity angle of the lepton $\theta_l$ and the angle between the leptonic and hadronic decay planes $\chi$. The corresponding terms in the differential decay rate are given by~\cite{Gustafson:2023lrz} 
\begin{align}
    &\frac{1}{K(q^2,M^2_{D\pi})}\frac{\mathrm{d}\Gamma^{(5)}}{\mathrm{d}q^2\,\mathrm{d}M^2_{D\pi}\,\mathrm{d}\chi\,\mathrm{d}\cos\theta\,\mathrm{d}\cos\theta_l}\Bigg|_\text{int}\nonumber\\ &=~\sin\theta\sin\theta_l\Bigg(\frac{1}{\sqrt{\lambda_B}}\mathrm{Re}(f_+ f^\ast_1)\cos\theta_l\cos\chi \nonumber\\ &+~\mathrm{Im}(f_+ g^\ast_1)\cos\theta_l\sin\chi\nonumber\\ &-~\mathrm{Re}(f_+ g^\ast_1)\cos\chi-~\frac{1}{\sqrt{\lambda_B}}\mathrm{Im}(f_+ f^\ast_1)\sin\chi\Bigg)\nonumber\\
    &+ \cos\theta\sin^2\theta_\ell (M_B^2 - M_{D\pi}^2)\frac{\mathrm{Re}(f_+ \mathcal{F}_{1,1}^\ast)}{\sqrt{q^2} \sqrt{\lambda_B}}~.
\end{align}
The prefactor is given by
\begin{align}
    K(q^2,M_{D\pi}^2) = \frac{G_F^2 |V_{cb}|^2}{4^8 \pi^6}\frac{\sqrt{q^2}\sqrt{\lambda_B} \lambda_{D\pi}}{M_B^3 M_{D\pi}^2}~.
\end{align}
While the forward--backward asymmetry of the $D$-meson is proportional to $\mathrm{Re}(f_+ \mathcal{F}_{1,1}^\ast)$, no counterpart containing the corresponding imaginary part exists. Thus, to extract the phase difference, knowledge about the form factors themselves is necessary. 
However, for the other combinations, $f_+f_1^*$
and $f_+ g_1^*$, we can measure both real and imaginary parts. This provides direct access
to the tangent of the phase difference, since the
norms of the form factors cancel in the ratios. Consequently, we will neglect the forward--backward asymmetry of the $D$-meson in the following.
\end{sloppypar}

The four terms can be projected out by integrating over $\cos\theta$, followed by antisymmetric integrations over $\cos\theta_l$ and $\chi$. We define the following operators to project on the various angular structures:
\begin{align}
    \mathcal{P}^{\cos\theta} &= \frac{2}{\pi}\int_{-1}^1\mathrm{d}\cos\theta~, \notag\\
    \mathcal{P}^{\cos\theta_l}_{++} &= \frac{2}{\pi}\int_{-1}^1\mathrm{d}\cos\theta_l~, \notag\\
    \mathcal{P}^{\cos\theta_l}_{-+} &= \frac{3}{2}\left(\int_{0}^1\mathrm{d}\cos\theta_l-\int_{-1}^0\mathrm{d}\cos\theta_l\right)~,\notag\\ \nonumber
    \mathcal{P}^{\chi}_{+--+} &= \frac{1}{4}\left(\int_{0}^{\pi/2}\mathrm{d}\chi
    - \int_{\pi/2}^{\pi}\mathrm{d}\chi\right. \notag\\
    & \qquad  \qquad \qquad \ \   - \left.\int_{\pi}^{3\pi/2}\mathrm{d}\chi + \int_{3\pi/2}^{2\pi}\mathrm{d}\chi\right)~, \notag\\
    \nonumber
    \mathcal{P}^{\chi}_{++--} &= \frac{1}{4}\left(\int_{0}^{\pi/2}\mathrm{d}\chi + \int_{\pi/2}^{\pi}\mathrm{d}\chi \right.\\
    & \qquad  \qquad \qquad \ \    \left.- \int_{\pi}^{3\pi/2}\mathrm{d}\chi - \int_{3\pi/2}^{2\pi}\mathrm{d}\chi\right) \ .
\end{align}
Thus, we obtain
\begin{align}
    \mathrm{Re}(f_+ f^\ast_1) &= \mathcal{P}^{\cos\theta}\mathcal{P}_{-+}^{\cos\theta_l}\mathcal{P}^{\chi}_{+--+} 
    \nonumber \\ & \qquad \qquad \qquad\times
    \frac{\sqrt{\lambda_B}}{K(q^2,M_{D\pi}^2)}\frac{\mathrm{d}\Gamma^{(5)}}{\mathrm{d}\Omega}~, \notag\\
    \mathrm{Im}(f_+ g^\ast_1) &= \mathcal{P}^{\cos\theta}\mathcal{P}_{-+}^{\cos\theta_l}\mathcal{P}^{\chi}_{++--}
     \nonumber \\ & \qquad \qquad \qquad\times
  \frac{1}{K(q^2,M_{D\pi}^2)}\frac{\mathrm{d}\Gamma^{(5)}}{\mathrm{d}\Omega}~, \notag\\
    \mathrm{Re}(f_+ g^\ast_1) &= -\mathcal{P}^{\cos\theta}\mathcal{P}_{++}^{\cos\theta_l}\mathcal{P}^{\chi}_{+--+}
     \nonumber \\ & \qquad \qquad \qquad\times
  \frac{1}{K(q^2,M_{D\pi}^2)}\frac{\mathrm{d}\Gamma^{(5)}}{\mathrm{d}\Omega}~, \notag\\
    \mathrm{Im}(f_+ f^\ast_1) &= -\mathcal{P}^{\cos\theta}\mathcal{P}_{++}^{\cos\theta_l}\mathcal{P}^{\chi}_{++--} 
     \nonumber \\ & \qquad \qquad \qquad\times
  \frac{\sqrt{\lambda_B}}{K(q^2,M_{D\pi}^2)}\frac{\mathrm{d}\Gamma^{(5)}}{\mathrm{d}\Omega}~.
\end{align}

The real and imaginary parts of the products of form factors are proportional to the cosine and sine of the phase difference, respectively. Consequently, in their ratio the form factors cancel and we are left with the tangent of the phase difference.
Since the phase shift of the $D\pi$ P wave, which is completely determined by the $D^*$, is known to be $\pi$ immediately above threshold, we get direct access to the $D\pi$ S-wave phase shift in a model-independent way.
Since two different P-wave form factors contribute, two ratios leading to the same tangent can be constructed, providing a valuable cross-check:
\begin{align}
\label{eq:tangent}
    \tan(\delta_0 - \delta_1) = \frac{\mathrm{Im}(f_+ f^\ast_1)}{\mathrm{Re}(f_+ f^\ast_1)}= \frac{\mathrm{Im}(f_+ g^\ast_1)}{\mathrm{Re}(f_+ g^\ast_1)}~.
\end{align}

\subsection{Measurements of asymmetries at Belle II}
$B$-factory experiments like Belle II are particularly well-suited for studies of semileptonic $B$-meson decays due to the distinct experimental setup of a precisely known initial state, coupled with an almost exclusive production of a $B\bar{B}$ pair through the $\Upsilon(4S)$ resonance. By exploiting this unique event topology, one of the bottom mesons can be fully reconstructed through purely hadronic decay chains, allowing for the kinematics of the remaining bottom meson to be inferred using conservation constraints, a method called hadronic tagging.

\begin{figure}
\begin{center}
\includegraphics[width=\linewidth]{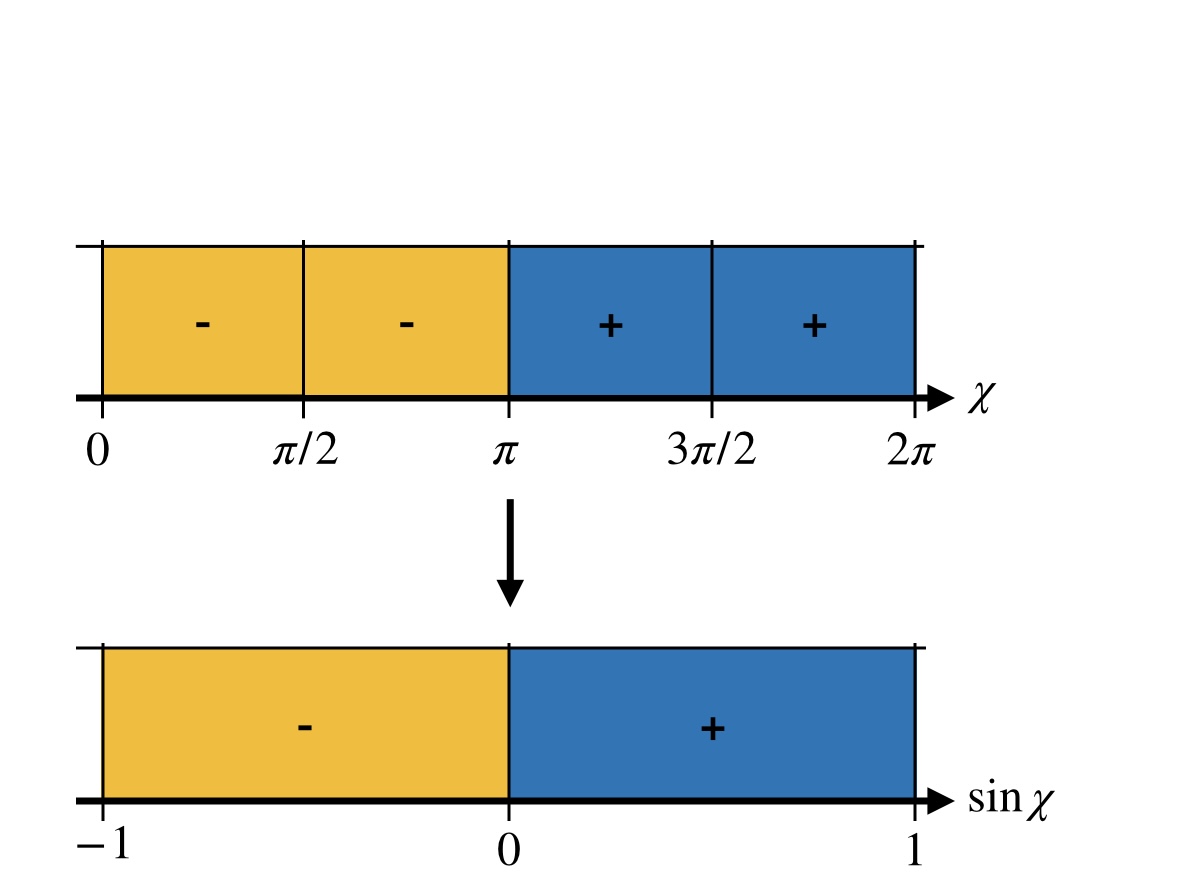}
\includegraphics[width=\linewidth]{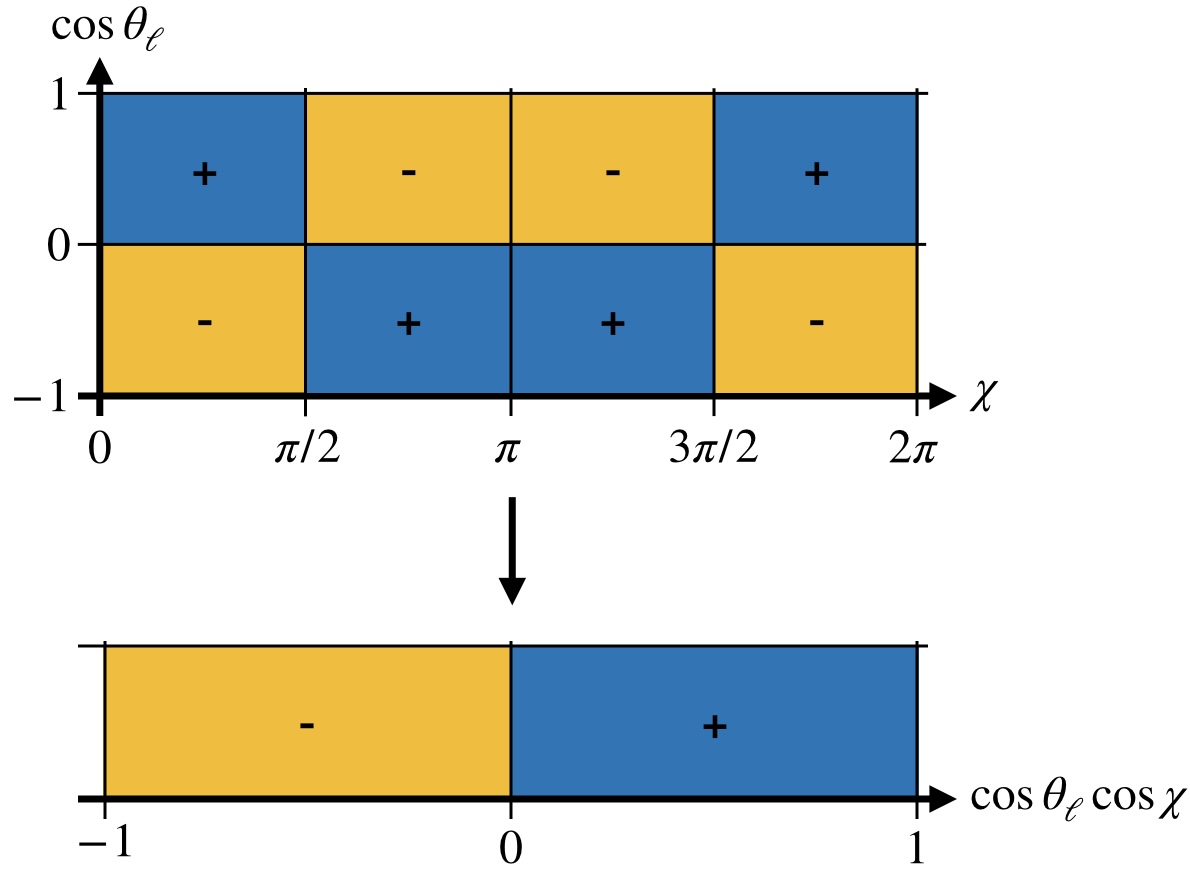}
\end{center}
\caption{Illustrative schematic of the equivalence between the two-dimensional asymmetry categories and the simplified one-dimensional categories. The asymmetry variables are defined as differences between yields of $+$ (blue)
and $-$ (yellow) regions of their respective angular variable(s). The upper observable is proportional to $\mathrm{Im}(f_+f_1^\ast)$ and the lower one to $\mathrm{Re}(f_+ f_1^\ast)$.}
\label{fig:obs:2d} 
\end{figure}

\begin{figure}
\begin{center}
\includegraphics[width=\linewidth]{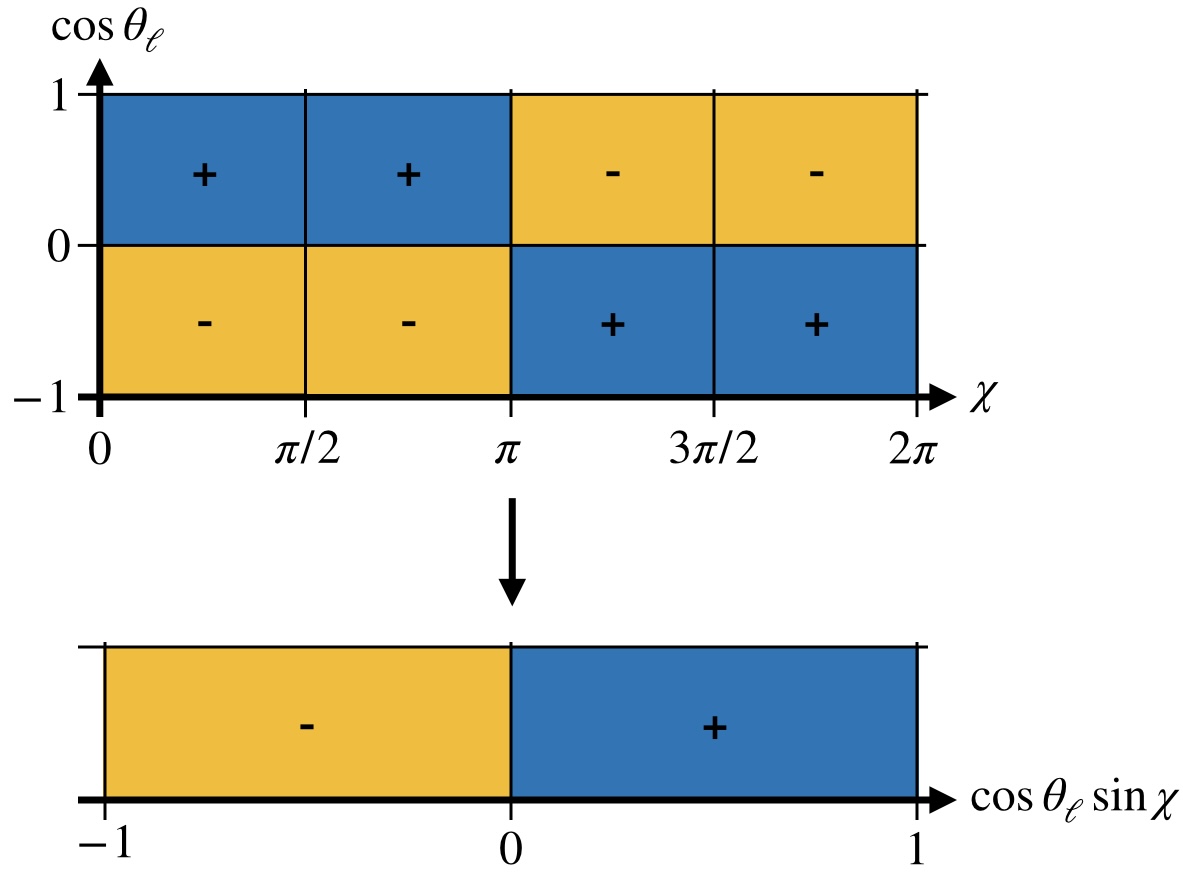}
\includegraphics[width=\linewidth]{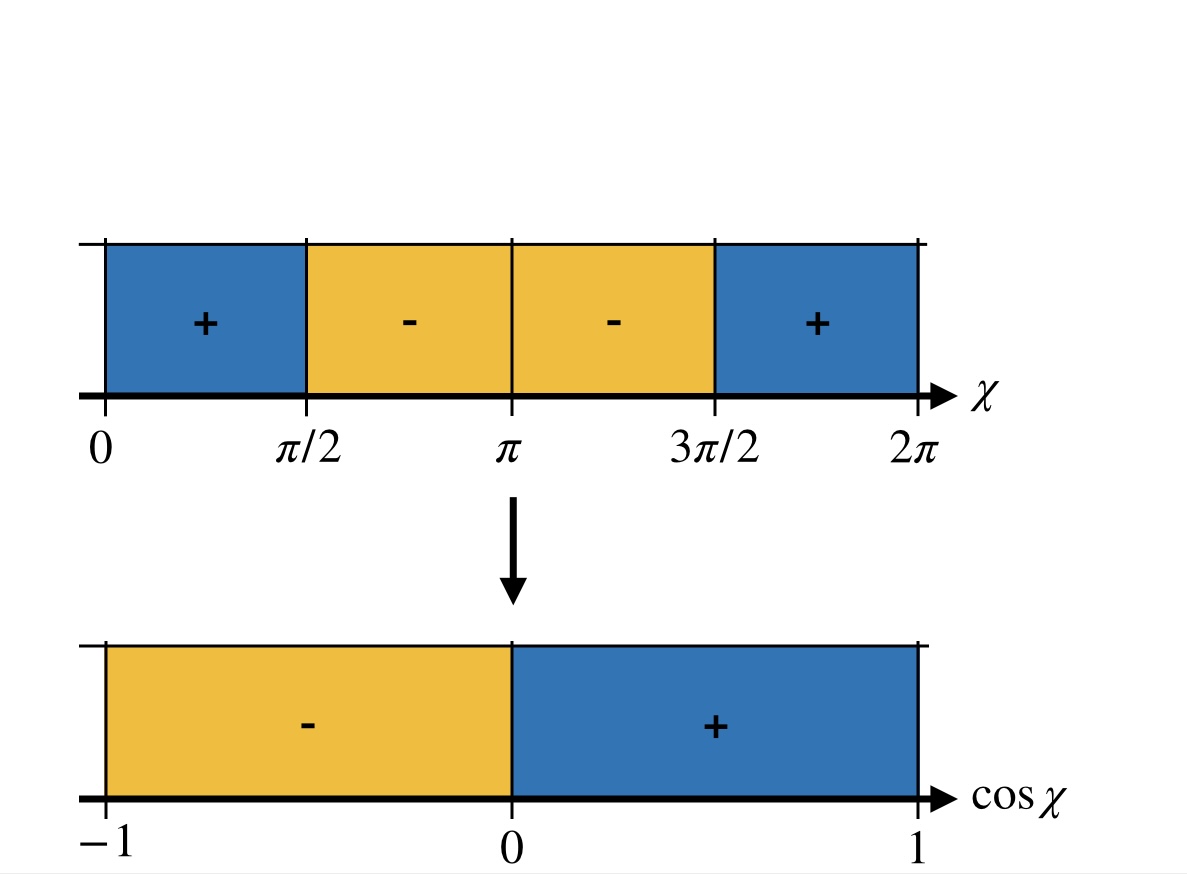}
\end{center}
\caption{Illustrative schematic of the equivalence between the one-dimensional asymmetry categories and the simplified one-dimensional categories. The asymmetry variables are defined as differences between yields of $+$ (blue)
and $-$ (yellow) regions of their respective angular variable(s). The upper observable is proportional to $\mathrm{Im}(f_+g_1^\ast)$ and the lower one to $\mathrm{Re}(f_+ g_1^\ast)$.}
\label{fig:obs:1d} 
\end{figure}

Belle and Belle II recently utilized this reconstruction technique to measure the angular-asymmetry observables of $B \rightarrow D^{\ast}\ell \nu $ decays~\cite{Belle:2023xgj,Belle-II:2023svm}. As the decay kinematics of $B \rightarrow D^{*}\ell \nu $ and $B \rightarrow D\pi\ell \nu $ decays are described by the same helicity angles, a similar strategy to the previous analyses can be employed to determine the angular observables defined in Sect.~\ref{sec:Obs} with the main difference that these variables are measured differentially in $M_{D\pi}$. Following the Belle II analysis, these asymmetries are redefined in terms of one-dimensional integrals
\begin{equation}
    \mathcal{A}_x(M_{D\pi}) \equiv \left(\frac{\mathrm{d}\Gamma}{\mathrm{d}M_{D\pi}}\right)^{-1} \left [ \int_{0}^{1} - \int_{-1}^{0}\right] \mathrm{d}x\frac{\mathrm{d}^2\Gamma}{\mathrm{d}M_{D\pi}\mathrm{d}x},
\end{equation}
with $x = \sin\chi$ for $\mathrm{Im}(f_+ f_1^\ast)$, $\cos\theta_{\ell}\cos\chi$ for $\mathrm{Re}(f_+f_1^\ast)$, $\cos\chi$ for $\mathrm{Re}(f_+g_1^\ast)$, and $\cos\theta_{\ell}\sin\chi$ for $\mathrm{Im}(f_+ f_1^\ast)$. The equivalence between the two-dimensional asymmetry categories and the one-dimensional categories is illustrated in Fig.~\ref{fig:obs:2d} and Fig.~\ref{fig:obs:1d}. Each of the angular asymmetries is determined by measuring the signal yields $N_{x}^{-}$ with $x\in[-1,0)$ and $N_{x}^{+}$ with $x\in[0,1]$ after accounting for detector resolution and acceptance effects. The asymmetries are then calculated as
\begin{equation}
    \mathcal{A}_{x}(M_{D\pi}) = \frac{N_{x}^{+}(M_{D\pi}) -N_{x}^{-}(M_{D\pi})}{N_{x}^{+}(M_{D\pi}) +N_{x}^{-}(M_{D\pi})} ~.
\end{equation}
This method has the advantage that various experimental uncertainties cancel in the asymmetries $\mathcal{A}_{x}$. Furthermore, measuring signal yields in $N_{x}^{-}$ and $N_{x}^{+}$ allows for higher reconstruction efficiencies, while reducing fluctuations due to limited statistics. Finally, to extract the tangent of the phase difference the appropriate ratios of the asymmetries $\mathcal{A}_{x}$ can be constructed as shown in Eq.~\eqref{eq:tangent}.

\section{Sensitivity study}\label{sec:Study}
In the following, we estimate the achievable precision on the pole location of the charmed scalar resonance $D_0^*$ and the S-wave $D\pi$ scattering length.
As a starting point, we take the measurement of the $M_{D\pi}$ spectrum in $B\rightarrow D\pi\ell\nu$ decays by the Belle experiment~\cite{Belle:2022yzd}.
This measurement uses the hadronic tagging method and measures the branching fraction and $M_{D\pi}$ distributions of $B\rightarrow D^{(\ast)}\pi^\pm\ell \nu_\ell$ decays. From this data, we extract the normalization of the S-wave component by updating the fit of Ref.~\cite{Gustafson:2023lrz} with the changes to the S- and P-wave contributions discussed in Sect.~\ref{sec:FFs}. While hadronic tagging would allow for precise measurements of angular distributions or the $q^2$ spectrum, no such measurement was performed in Ref.~\cite{Belle:2022yzd}.

Next, based on the fit results, we extrapolate the number of events in the $B^+ \rightarrow D^- \pi^+ \ell^+ \nu_\ell$ channel\footnote{We do not consider the $B^0 \rightarrow \bar{D}^0 \pi^- \ell^+ \nu_\ell$ channel here, as it has larger uncertainties in the Belle analysis~\cite{Belle:2022yzd} and the $D^\ast$ resonance is above threshold, possibly leading to a significant change in reconstruction efficiency, thus inhibiting our extrapolation.} obtained in Ref.~\cite{Belle:2022yzd} in the region of $M_{D\pi}\in [2.05,2.3]~\text{GeV}$ to $M_{D\pi}\in [(M_{D^-}+M_{\pi^+}),2.3~\text{GeV}]$, leading to $400\pm 23$ events. In the following, we scale this number, corresponding to an integrated luminosity of $772~\mathrm{fb}^{-1}$ at the $\Upsilon(4S)$ resonance, to different luminosity scenarios. We will study two scenarios for which data have already been recorded: the current Belle II dataset, as well as the combined Belle and Belle II dataset.
Furthermore, we make projections for $2~\mathrm{ab}^{-1}$, $5~\mathrm{ab}^{-1}$, and $10~\mathrm{ab}^{-1}$ of data. Belle II is expected to reach these target integrated luminosities by 2027, 2030, and 2032, respectively, according to the current projection plan~\cite{B2Luminosity}.

To generate the pseudo-data we sample events from the five-fold differential rate, based on the central values for the form factors. Next, we compute the $N_x^+$ and $N_x^-$ for each observable in six bins of $M_{D\pi}$ between $2.0$ and $2.3$ GeV and determine the statistical uncertainties through bootstrapping. We then compute the asymmetries $\mathcal{A}_x$ in each $M_{D\pi}$ bin and take the respective ratios.

Finally, we subtract the P-wave phase $\delta_1$, based on the Chew--Mandelstam improved Breit--Wigner lineshape.
From the corrected data, we can extract the pole location of the $D_0^\ast$, as well as the isospin-$1/2$ S-wave scattering length.

\subsection{Pole location}

To extract the pole location from the pseudo-data (as well as later for the real data), we parameterize the scattering amplitude in a modified $K$-matrix formalism, taking into account the correct chiral behavior near threshold (see discussion in Sect.~\ref{subsec:swave}):
\begin{align}
    T(s) &= \frac{K(s)}{1 - \Sigma_0(s + i\epsilon) K(s)}~,\\
    K(s) &= \frac{E_\pi (g_0 + g_1 s)^2}{s - M_R^2} + E_\pi g_2~.
\end{align}
Here, the phase space factor in the Chew--Mandelstam representation, $\Sigma_0$, is given in Eq.~\eqref{eq::sig0}.

For each scenario, we find the minimum of the likelihood using the Migrad algorithm as implemented in \texttt{iminuit}~\cite{James:1975dr,iminuit}. In the next step, we compute confidence regions with the Minos algorithm and determine the pole location on the borders of these regions.

\begin{figure}
\begin{center}
\includegraphics[width=\linewidth]{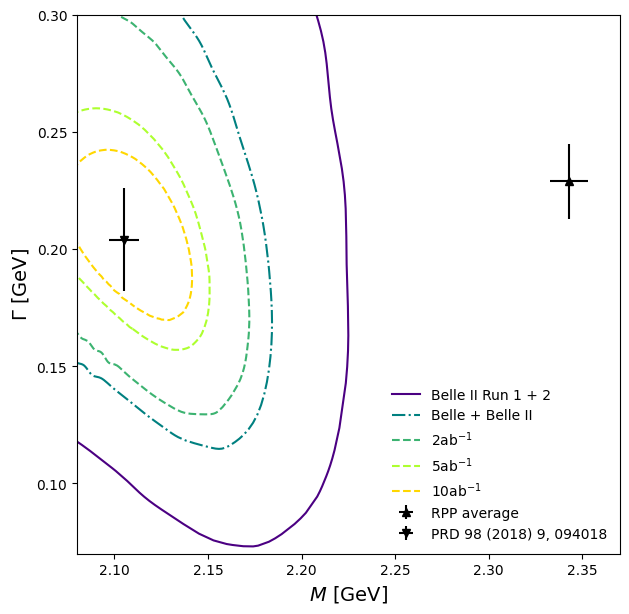}
\end{center}
\caption{The pole location $M + i\Gamma/2$
extracted from the pseudo-data
generated from the amplitude of
Ref.~\cite{Liu:2012zya}
in the mass versus width plane. For the five different scenarios we show the $68\%$ confidence contours. In addition, we show the pole location from Ref.~\cite{Du:2017zvv}, as well as the average from the RPP~\cite{ParticleDataGroup:2024cfk}.}
\label{fig:mass:width} 
\end{figure}

The results of the minimal scenario with $g_1 = g_2 = 0$ for the $1\sigma$ region are shown in Fig.~\ref{fig:mass:width}. Even with the current Belle II dataset alone, the current Review of Particle Physics (RPP) average~\cite{ParticleDataGroup:2024cfk} for the $D_0^*(2300)$ resonance parameters could be ruled out at the $2.2\sigma$ level. A combined analysis of the existing Belle and Belle II datasets could rule out the RPP average by more than $3\sigma$, while $2~\text{ab}^{-1}$ would be sufficient for $5\sigma$. While even with $10~\text{ab}^{-1}$ of data, the pole location calculated in Ref.~\cite{Du:2017zvv} remains more precise than our projected experimental accuracy, any experimental improvement over the assumed efficiencies will make the direct extraction competitive with the existing theoretical determination.

Allowing for either $g_1$ or $g_2$ to be non-zero in the fit does not change the outcome significantly: with the current Belle II dataset, the RPP value could still be ruled out at the $2\sigma$ level\footnote{Allowing for $g_1$ and $g_2$ to be non-zero at the same time leads to large correlations between the two, but no change in outcome.}. However, with increasing statistics, uncertainties due to the truncation order of the $K$-matrix parameterization need to be assessed.

\subsection{Scattering length}
\begin{sloppypar}
In a non-relativistic normalization the $T$-matrix $T_{\rm NR}$ for elastic $D\pi$ scattering may be expressed as
\begin{equation}
 T_{\rm NR \, 0}^{(1/2)} = -\frac{2\pi}{\mu}
 \frac{1}{q_1\cot\delta_0^{(1/2)}-iq_1} \ ,
 \label{eq:deltadef}
\end{equation}
with $q_1$  for the c.m.\ momentum of the $D\pi$ system, as before.
Traditionally, the scattering length, $a_0^{(1/2)}$,
and the effective range, $r_0^{(1/2)}$, are defined through the effective-range expansion (ERE):
\begin{align}
    q_1\cot\delta_0^{(1/2)} = \frac{1}{a_0^{(1/2)}} + \frac{1}{2}r_0^{(1/2)} q_1^2 +\mathcal{O}(q_1^4)~.\label{eq::ere}
\end{align}
However, the radius of convergence of the ERE is restricted by the location of the nearest singularity.
As a consequence of the approximate chiral symmetry of QCD, the scattering amplitude has an Adler zero right below the threshold (see discussion in Sect.~\ref{subsec:swave}),
 which, according to Eq.~(\ref{eq:deltadef}),
  translates into a pole of $q_1\cot\delta_0^{(1/2)}$. Thus, for a reliable extraction of the scattering length
from a fit to data, the ERE needs to be modified
and we propose to employ a modified ERE with a significantly enhanced radius of convergence:
\begin{align}
    q_1\cot\delta_0^{(1/2)} = \frac{c_{-1}}{E_\pi}+c_0+
    c_1 E_\pi + \mathcal{O}(E_\pi^2)
  \ .\label{eq:eff2}
\end{align}
Once the parameters $c_i$ are determined 
from the fit to the phase shifts, the scattering length
can be extracted from the threshold value of 
Eq.~(\ref{eq:eff2}),
namely
\begin{equation}
    a_0^{(1/2)}=\frac{M_\pi}{c_{-1}+c_0 M_\pi +
    c_1 M_\pi^2} \ .
\end{equation}
The values of $a_0^{(1/2)}$ extracted from the pseudo-data, as well as lower limits at $90\%$ confidence level for the five scenarios, are given in Table~\ref{tab:scatteringlength}.
\end{sloppypar}

Due to large correlations between the different coefficients, a direct measurement of the scattering length proves challenging, especially at lower statistics. In particular, for the existing Belle II scenario, only a lower limit on the scattering length can be set. While a precision measurement of the scattering length requires significantly more data,\footnote{Note, that the $K\rightarrow\pi\pi\ell\nu$ were conducted with two orders of magnitude more signal events than the $10~\text{ab}^{-1}$ scenario~\cite{NA482:2010dug}.} almost all studied scenarios can set lower limits at the $90\%$ confidence level challenging the ALICE results.

\begin{table}
\caption{\label{tab:scatteringlength} Extracted value of the scattering length, as well as $90\%$ confidence level lower limits in the sensitivity study for the five scenarios.}
\begin{tabular}{Sc | Sc | Sc}
Scenario & $a_0^{(1/2)}$ & $a_0^{(1/2)}$ at $90\%$ CL\\
\hline
Belle II & $0.45^{+\infty}_{-0.39}~\text{fm}$ & $> 0.03~\text{fm}$ \\
Belle + Belle II & $0.44^{+16.02}_{-0.34}~\text{fm}$ & $> 0.04~\text{fm}$ \\
$2~\text{ab}^{-1}$ & $0.45^{+2.96}_{-0.31}~\text{fm}$ & $> 0.05~\text{fm}$ \\
$5~\text{ab}^{-1}$ & $0.45^{+0.92}_{-0.24}~\text{fm}$ & $> 0.11~\text{fm}$ \\
$10~\text{ab}^{-1}$ & $0.45^{+0.51}_{-0.20}~\text{fm}$ & $> 0.15~\text{fm}$
\end{tabular}
\end{table}

\begin{figure}[htbp]
\begin{center}
\includegraphics[width=\linewidth]{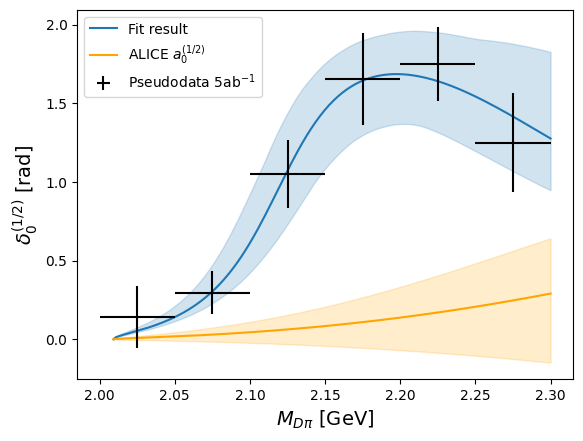}
\end{center}
\caption{Example scattering phase fit for the $5~\text{ab}^{-1}$ scenario. The pseudo-data is shown together with the result of the fit to the parameterization in Eq.~\eqref{eq:eff2} and a curve obtained from ALICE' measurement of $a_0^{(1/2)}$, but neglecting the higher orders in the effective range expansion ($r_i^{(1/2)} = 0$ in Eq.~\eqref{eq::ere}).}
\label{fig:a0} 
\end{figure}

In Fig.~\ref{fig:a0}, we show our fit result for the $5~\text{ab}^{-1}$ scenario, the corresponding simulated data, and the expectation based on ALICE' measurement of $a_0^{(1/2)}$. For the ALICE curve we set higher orders in the ERE to zero, but include the Adler zero.
It is clearly visible that near the threshold, where the approximation for the ALICE curve is valid, the slopes differ significantly.

\section{Conclusion and outlook}\label{sec:Conc}
\begin{sloppypar}
In this work we proposed observables allowing for the direct extraction of the $D\pi$ S-wave scattering phase from semileptonic $B\rightarrow D\pi\ell\nu$ decays. We have demonstrated that with the existing Belle and Belle II datasets, a meaningful determination of the pole location of the lightest scalar charmed meson could be achieved. In addition, with continued data taking of the Belle II experiment, also the $D\pi$ S-wave scattering length can be measured to sufficient precision to challenge recent measurements reported by the ALICE collaboration.
\end{sloppypar}

The sensitivity estimates provided here are all based on the hadronic tagging method, which provides excellent angular resolution, but is statistically limited by the tagging efficiency. However, there is significant room for improvement beyond the sensitivities obtained here. First, $B^0\rightarrow \bar{D}^0\pi^-\ell^+\nu_\ell$ decays can be combined with the $B^+\rightarrow D^-\pi^+\ell^-\nu$ decays studied here. Second, our study is based on efficiencies obtained by the Belle collaboration, while the Belle II detector provides improved particle identification.

Given the promising results obtained here, it would be worth studying the use of the semileptonic tagging method or performing an untagged measurement at Belle II, providing larger reconstruction efficiencies at the cost of increased backgrounds and more complicated reconstruction of the required angles. Encouragingly, the Belle II Collaboration already performed a measurement of the $\cos\theta_\ell$ and $\chi$ distributions in untagged $B^0\rightarrow D^{\ast-}(\rightarrow \bar{D}^0 \pi^-)\ell^+\nu_\ell$ decays.

On the theoretical side, it is worth extending our analysis of asymmetry observables by including the D wave, thus allowing for the extraction of the difference of S- and D-wave phase shifts. Since the $D\pi$ D-wave is dominated by the Breit--Wigner-like $D_2^\ast(2460)$ resonance, it might be possible to observe the S-wave phase motion near the $D\eta$ and $D_s \bar K$ thresholds, which is expected to show nontrivial structure in UChPT~\cite{Du:2017zvv}.

\begin{acknowledgements}
We thank Ruth Van de Water for collaboration and discussion during the initial stages of this project, as well as Markus Prim for comments on the manuscript.
This research is supported in part by the Swiss National Science Foundation (SNF) under contract 200021-212729
and by the MKW NRW
under funding code NW21-024-A. RvT is supported by the German Research Foundation (DFG) Walter-Benjamin Grant No.~545582477.
FKG is supported in part by the National Natural Science Foundation of China under Grants No.~12361141819, No.~12125507, and No.~12447101; and by the Chinese Academy of Sciences (CAS) under Grant No.~YSBR-101. 
CH\ thanks the CAS President's International Fellowship Initiative (PIFI) under Grant No.\ 2025PD0087 for partial support.
\end{acknowledgements}

\bibliographystyle{utphysmod}
\bibliography{refs}

\end{document}